\documentclass[useAMS,usenatbib,usegraphicx]{mn2e}

\newcommand\suzaku{{\it Suzaku}}

\newcommand\rxte{{\it RXTE}}

\newcommand\swift{{\it Swift}}
\newcommand\maxi{{\it MAXI}}
\newcommand\integral{{\it INTEGRAL}}
\newcommand\ftools{{\it FTOOLS}}

\newcommand\kev{{\rm~keV }}
\newcommand\ev{{\rm~eV}}
\newcommand\bcol{}
\newcommand\kms{\ifmmode {\rm~km\ s}^{-1} \else ~km s$^{-1}$\fi}
\newcommand\Hunit{\ifmmode {\rm~km\ s}^{-1}\ {\rm Mpc}^{-1}
        \else ~km s$^{-1}$ Mpc$^{-1}$\fi}
\newcommand\ctssec{\ifmmode {\rm~count\ s}^{-1} \else ~count s$^{-1}$\fi}
\newcommand\ergsec{\ifmmode {\rm~erg\ s}^{-1} \else
        ~erg s$^{-1}$\fi}
\newcommand\funit{\ifmmode {\rm~erg\ s}^{-1}\;{\rm cm}^{-2} \else
        ~ergs s$^{-1}$ cm$^{-2}$\fi}
\newcommand\phflux{\ifmmode {\rm~photon\ s}^{-1}\;{\rm cm}^{-2}
        \else   ~photon s$^{-1}$ cm$^{-2}$\fi}
\newcommand\efluxA{\ifmmode {\rm~erg\ s}^{-1}\;{\rm cm}^{-2}\;{\rm
        \AA}^{-1} \else ~erg s$^{-1}$ cm$^{-2}$ \AA$^{-1}$\fi}
\newcommand\efluxHz{\ifmmode {\rm~erg\ s}^{-1}\;{\rm cm}^{-2}\;{\rm
        Hz}^{-1} \else ~erg s$^{-1}$ cm$^{-2}$ Hz$^{-1}$\fi}
\newcommand\cc{\ifmmode {\rm~cm}^{-3} \else cm$^{-3}$\fi}
\newcommand\FWHM{\ifmmode {\rm~FWHM} \else ${\rm~FWHM}$\fi}
\newcommand\Msun{\ifmmode M_{\odot} \else $M_{\odot}$\fi}
\newcommand\Lsun{\ifmmode L_{\odot} \else $L_{\odot}$\fi}

\newcommand\hbeta{\ifmmode {\rm H}\beta \else H$\beta$\fi}
\newcommand\Kalpha{\ifmmode {\rm K}\alpha \else K$\alpha$\fi}
\newcommand\nh{\ifmmode N_{\rm H} \else N$_{\rm H}$\fi}

\title[Cyclotron line variation in 4U0115+63]{ Variations in the Cyclotron Resonant
  Scattering Features during 2011 outburst of 4U~0115+63}
\author[Iyer et~al.]{N. Iyer$^{1,2}$\thanks{E-mail: nirmal@physics.iisc.ernet.in}, 
D.Mukherjee$^{3}$\thanks{E-mail: dipanjan.mukherjee@anu.edu.au},
G.C Dewangan$^{4}$\thanks{E-mail: gulabd@iucaa.ernet.in},
D. Bhattacharya$^{4}$\thanks{Email: dipankar@iucaa.ernet.in} and
S.Seetha$^{5}$\thanks{E-mail: seetha@isro.gov.in} \\
$^{1}$ISRO Satellite Centre, Bangalore , India\\
$^{2}$Indian Institute of Science, Bangalore, India \\
$^{3}$ Research School of Astronomy and Astrophysics, The Australian National University, Canberra, ACT 2611, Australia \\
$^{4}$Inter-University Centre for Astronomy and Astrophysics, Pune , India\\
$^{5}$Space Science Office, ISRO HQ, Bangalore , India\\
}

\begin{document}
\date{Accepted **. Received **; in original form **}
\pagerange{\pageref{firstpage}--\pageref{lastpage}} \pubyear{201x}

\maketitle

\label{firstpage}

\begin{abstract}
  We study the variations in the Cyclotron Resonant Scattering Feature 
(CRSF) during 2011 outburst of the high mass X-ray binary 4U~0115+63 using observations 
performed with \suzaku{}, \rxte{}, \swift{} and \integral{} satellites. The
wide-band 
spectral data with low energy coverage allowed us to characterize the broadband 
continuum and detect the CRSFs. We find that the broadband continuum 
is adequately described by a combination of a low temperature ($kT\sim 0.8\kev$) blackbody 
and a power-law with high energy cutoff ($E_{cut}\sim 5.4\kev$) without the need for a 
broad Gaussian at $\sim 10\kev$ as used in some earlier studies. 
Though winds from the companion can affect the emission from the neutron 
star at low energies ($<$ 3 keV), the blackbody component shows a 
significant presence in our continuum model.
We report evidence for the possible presence of two 
independent sets of CRSFs with fundamentals at $\sim 11\kev$ and $\sim 15\kev$.
These two sets of CRSFs could arise from spatially distinct emitting regions.  
We also find evidence for variations in the line equivalent widths, with the $11\kev$ 
CRSF weakening and the $15\kev$ line strengthening with decreasing luminosity.
Finally, we propose that the reason for the earlier observed anti-correlation of line
energy with luminosity could be due to modelling of these two independent line
sets ($\sim 11\kev$ and $\sim 15\kev$) as a single CRSF.

\end{abstract}


\section{Introduction}\label{sec:int}
4U~0115+63 is a high mass X-ray binary system, first discovered in the
UHURU satellite's sky survey \citep{gia72,for78}, with more than 15 subsequent 
outbursts recorded till date \citep{bol13}. The system consists of a pulsating neutron 
star with spin period $\sim$ 3.61s \citep{com78} and a B0.2Ve main sequence star
\citep{joh78}, with an orbital period 
of $\sim$ 24.3 days \citep{rap78}. The distance to this binary system has been 
estimated to be $\sim 7{\rm~kpc}$ \citep{neg01}. 
The source exhibits luminous Type II X-ray outburst during which multiple
cyclotron resonance scattering features (CRSF) have been observed in the
X-ray spectrum, with 5 detected harmonics \citep{san99,fer09}.
CRSF  are caused by scattering of 
X-ray photons from electrons in the accreting plasma channeled by the NS magnetic field.
The energy at which these lines occur is given as ${E_{cyc} = 11.6B_{12}\,\times\,(1+z)^{-1}}$\kev
\citep{cob02}. Here $B_{12}$ is the local 
magnetic field (in units of $10^{12}{\rm~Gauss}$) and $z$ is the gravitational 
redshift in line energy. Thus cyclotron lines give us a direct probe of the 
local magnetic field near the scattering regions.

Cyclotron line parameters of many sources are found to vary with the phase of
rotation \citep[see][for a review]{hei04}, and the varying luminosity 
of the outburst \citep{bec12}. However, the variation in energy of the 
fundamental CRSF of 4U~0115+63 with luminosity has been the source of 
some debate. \citet{nak06,tsy07,li12} find an 
anti-correlation between the line energy and luminosity,  whereas \citet{mul13} 
find this anti-correlation to be an artifact of the continuum spectral 
modelling. \citet{bol13} have pointed out the cause of this reported 
dichotomy to be due to the use of a broad Gaussian like emission feature 
to model the continuum in some of the works.

In this paper, we study multiple observations during the 2011 outburst of 
4U~0115+63  performed with \suzaku{}, \rxte{}, \swift{} and \integral{} 
satellites, providing us with a wide-band coverage from $0.5\kev$ to
$\sim 60\kev$ with high signal-to-noise. As shall be seen in subsequent 
sections, this availability of wide bandwidth data is very important 
to correctly model both the continuum and the CRSF in the source spectrum. 
From the results of our spectral analysis, we find evidence for two sets of
cyclotron lines 
whose parameters vary with source luminosity. 
In the following sections, we describe the observations and  
reduction of data that we used in \S \ref{ss:obs},
 spectral analysis of these data-sets in \S \ref{ss:mod}, 
and the inferences and  possible implications in \S \ref{ss:con}

\section{Observations and Data Reduction}\label{ss:obs}
\begin{figure}
  \centering
  \includegraphics[width=9cm]{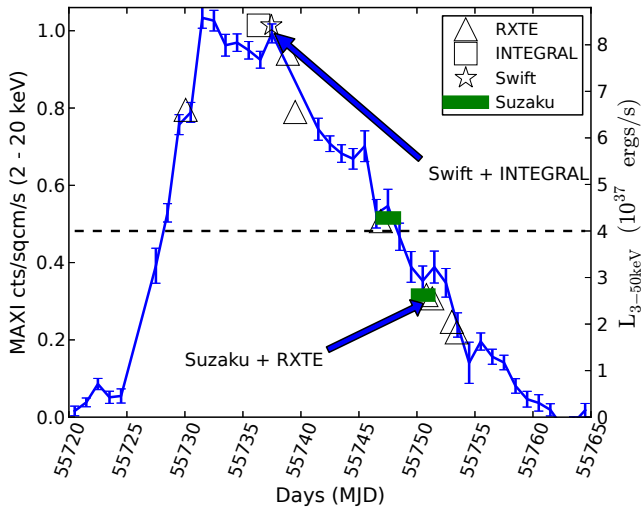}
  \caption{Variations in the count rate and luminosity during the 2011 outburst of 4U~0115+63. 
    The continuous line represents the observed count rate measured with the \maxi{} sky monitor. 
	The symbols represent the pointed observations and the corresponding  $3 -50\kev$ luminosities
	derived from our spectral analysis. The two simultaneous wide-band 
	observations are annotated separately. The horizontal line
	shows the luminosity level($4\times 10^{37}{\rm ergs\, s^{-1}}$) around 
	which the previous observations \citep{li12} have found a sharp change in
	fundamental CRSF energy. } 
  \label{fig:obs}
\end{figure}

In all, we analysed data obtained by different X-ray observatories over the 15 days of the 2011 outburst.
Table \ref{tab:obs} lists all the \rxte{} and \suzaku{} observations which 
were available in the archive and a \swift{} and an \integral{}
observation, made when the source was near its peak luminosity. Figure~\ref{fig:obs}
shows the variations in the count rates as measured with \maxi{} sky monitor
during the 2011 outburst. The pointed observations 
performed at different luminosity levels are also marked in this figure. 
The reduction of data-sets from each of these satellites is explained 
in the following sub-sections.

\begin{table}
  \centering
  \caption{A summary of pointed observations during the 2011 outburst of 4U~0115+63. 
  }
  \begin{tabular}{c|c|c|c|l}
	\hline
	\textbf{MJD} & \textbf{Instrument} & \textbf{ObsId} & \textbf{MAXI$^a$} & \textbf{Exposure$^b$}\\ \hline
	\texttt{55730.06} & \rxte{} & 96032-01-01-00 & 0.788 & 6.53 \\
	\texttt{55736.34} & \swift{} & 00031172010 & 0.925 & 6.69\\
	\texttt{55736.34} & \integral{} & 106100650010 & 0.925 & 2.13\\
	\texttt{55738.87} & \rxte{} & 96032-01-02-00 &  0.993 & 4.52\\
	\texttt{55739.51} & \rxte{} & 96032-01-02-01 & 0.745 & 3.69\\
	\texttt{55743.91} & \rxte{} & 96032-01-03-00 & 0.682 & 5.02\\
	\texttt{55746.90} & \rxte{} & 96032-01-03-02 & 0.526 & 9.04\\
	\texttt{55747.53} & \suzaku{} & 406048010 & 0.526 & 24.27\\
	\texttt{55750.82} & \rxte{} & 96032-01-04-00 & 0.352 & 16.91 \\
	\texttt{55750.82} & \suzaku{} & 406049010 & 0.352 & 81.68\\
	\texttt{55751.32} & \rxte{} & 96032-01-04-02 & 0.388 & 2.08 \\ 
	\texttt{55753.00} & \rxte{} & 96032-01-04-03 & 0.248 & 1.85 \\ 
	\texttt{55753.42} & \rxte{} & 96032-01-04-04 & 0.238 & 0.73 \\ \hline
  \end{tabular}
    {\sf $^a$MAXI rate (in ${\rm counts~cm^{-2}~s^{-1}}$) is of nearest 
     observation. \\
      $^b$Exposure time is in kilo-seconds.}
  \label{tab:obs}
\end{table}

\subsection{\rxte{} Observations}
Among the \rxte{} observations, we used all but two of them. Obs-Id
96032-01-04-03 and 96032-01-04-04 were made when source flux was very low,
and with short exposure times (1856s and 736s respectively).
Due to poor count statistics, we were unable to constrain the CRSF parameters for these 
observations, making them unsuitable for the present work. \rxte{} spectra were 
obtained from the raw data files using \ftools{} from 
{\tt HEASoft v 6.15.1}. We used data from both the {\it HEXTE} and the {\it PCA} detectors.
PCU2 Science Array data were used for generating {\it PCA} spectrum, and Cluster A
Science Array data were used for {\it HEXTE} spectrum. {\it HEXTE} background was obtained
from Cluster B, and dead-time corrections were applied to both the source
and background {\it HEXTE} spectra. For {\it PCA}, we found the dead-time to be a maximum of 
5\%, which lowered its flux by about 5\%. We did separately correct for
this. 
Data grouping and usage of systematic errors
for the \rxte{} spectra are shown in Table \ref{tab:grp}. In all, the \rxte{}
provided usable data covering the 3 \kev to 50 \kev band.

\subsection{\suzaku{} Observations}
\suzaku{} had two long duration observations made during the course of the
2011 outburst. For reduction of \suzaku{} data, we used \ftools{} from
{\tt HEASoft 6.15.1} with {\tt CALDB} updated till June 2014. 
We used data from the {\it XIS} ( 0.6 \kev -- 10 \kev) and {\it PIN}
( 15 \kev -- 60 \kev). The {\it GSO} data had very low SNR and hence was not used.
Although the {\it PIN} does collect data from 12 \kev
onwards, we discarded the data from 12 \kev to 15 \kev due to 
high uncertainty in {\it PIN} background\footnote{see Suzaku Data Reduction ABC guide (version 4.0)}. 
{\it XIS} data above 10 \kev and {\it PIN} data above 60 \kev were too noisy 
to be of use for our analysis. As a result we were not able to 
effectively cover the energy range {(10 \kev -- 15 \kev)}, which is 
important to model the fundamental cyclotron line in this system. To overcome this, we used
simultaneously taken \rxte{} observations in conjunction with our
\suzaku{} data-sets. Since Obs-Id 406049010 (\suzaku{}) and Obs-Id
96032-01-04-00 (\rxte{}) had some overlap, we used the portion where
overlap existed. This was done by creating a GTI which covered the common 
interval between the \rxte{} and \suzaku{} GTI files. However Obs-Id 
406048010 (\suzaku{}) had no overlap with any other \rxte{} observation.
Thus, we did not use this data-set.

The {\it XIS} data were taken in the $1/4$ windowed mode (of size 256 pixels) at 
normal clocking speeds to reduce the effect of pile-up.
We found significant pile-up upto a maximum level of $\sim17\%$ 
in the central regions and corrected for it using the recipe of
John Davis\footnote{http://space.mit.edu/CXC/software/suzaku/}.
This was achieved by rejecting a central 
circular region of size $\sim25\arcsec$ from the {\it XIS} image for which the computed Pile-up 
percentage was greater than $6\%$. After extraction of data from individual 
{\it XIS} chips, the {\it XIS} 0 and 3 data were combined using the \texttt{addascaspec} tool.
This resulted in two sets of spectral files; one from the back illuminated (BI)
CCD  and the other from the combined front illuminated (FI) CCDs. {\it XIS} data below 0.8\kev were 
rejected owing to discrepancies  between the FI and BI spectra at a $\sim 3\sigma$ level.
A mismatch at a $\sim 2\sigma$ level between the two spectra around the Si K-edge was also
noticed, but we did not discard this part of the spectrum as it was
important for fitting our continuum model. This led to poorer
values of the $\chi^2$ statistic that we use for fitting the spectral data.

\begin{figure*}
  \centering
    \begin{minipage}{0.45\textwidth}
  	  \includegraphics[height=\textwidth,angle=270]{finneconspec.ps}
    \end{minipage}
    \begin{minipage}{0.45\textwidth}
      \includegraphics[width=\textwidth]{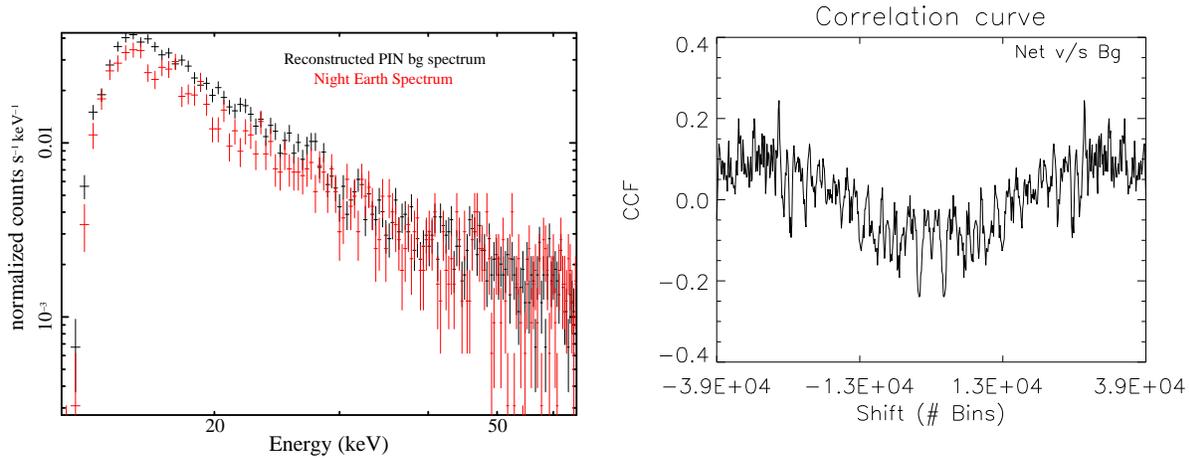}
    \end{minipage}
    \caption{A comparison of the estimated tuned {\it PIN} background and the night-earth 
	  {\it PIN} data. The left panel compares the spectral data extracted from reconstructed {\it PIN} 
	  background and the Night Earth data. See online edition for a color print of this figure. 
	  Right panel show the cross-correlation between the net (source - background) and background lightcurves. See 
	  text for details. }
  \label{fig:nebg}
\end{figure*}

The {\it PIN} data were extracted using the tool \texttt{hxdpinxbpi}.
In the joint analysis of \suzaku{} and \rxte{}, we required a cross-normalisation 
of $\sim$ 1.57, between the \suzaku{} {\it XIS} and {\it PIN}, much larger than the 
recommended value of $1.16$ given in the Suzaku Data Reduction ABC guide 
(version 4.0). Therefore, we investigated this issue 
in some detail. To verify the {\it PIN} data, we extracted individual 64 {\it PIN} 
count spectrum as explained in the \suzaku{} note on estimating {\it PIN} noise 
\footnote{http://heasarc.gsfc.nasa.gov/docs/suzaku/analysis/pinnoise.html}, 
and found no discrepancies. We also extracted night-earth data for the {\it PIN}
observations to compare and see if the estimated backgrounds were correct. 
Fig.~\ref{fig:nebg} compares the night-earth and the ``tuned'' background spectral 
data. The two background spectra are similar in shape and in flux. 

\begin{figure*}
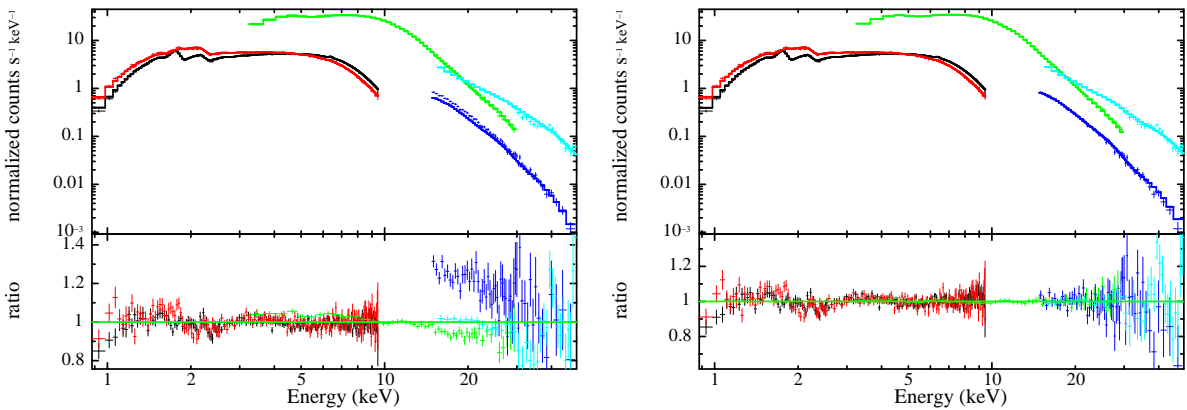

  \centering
  \includegraphics[height=0.45\textwidth,angle=270]{finrat116.ps}
  \includegraphics[height=0.45\textwidth,angle=270]{finrat157.ps}
  \caption{The spectral data ({\it XIS} data in red and black, {\it PIN} 
  data in dark blue, {\it PCA} data in green and {\it HEXTE} in light blue color) and the ratio plots
  for best-fitting {\tt cutoffpl} model. These plots were made 
  with the {\it XIS}-to-{\it PIN} cross-normalisation fixed at $1.16$ (left panel) and  
  with the best-fit value of $1.57$ (right panel). The offset is seen in the ratio plot in left panel when using a 
  cross-normalisation value of 1.16. See beginning of section 3 for comments on
  this figure and the online edition for a color print of this figure.  
 }  
  \label{fig:rat}
\end{figure*}

We further examined the validity  of the background subtraction by comparing 
the  net (source minus background) and background light curves. The lack of correlation 
amongst the two indicates that the {\it PIN} background was estimated correctly. 
Finally we checked if the {\it XIS} 
spectra were properly area corrected. The {\it XIS} data were taken in the 1/4 window
mode with normal clocking, thereby making a rectangular source footprint.
By choosing both rectangular and circular source regions for extraction 
we found no difference in the resulting spectrum. The difference in the fit-quality
as a result of using the recommended and the best-fit cross-normalisation factor, 
$1.16$ and $1.57$ respectively, is shown in 
Figure \ref{fig:rat}. As seen in the figure, the discrepancy in the
ratio plot for \suzaku{} {\it PIN} (seen as a set offset points in the ratio
plot of the left panel) vanishes when we use the
best fit cross normalisation factor. We have thus used a cross-normalisation 
factor of 1.57 for all our subsequent models and fits (also see Sect.~3).

\subsection{\swift{} and \integral{} observation}
We selected this particular observation from multiple \swift{} observations
of the source, because it was simultaneously taken with an \integral{}
observation (thereby enabling wide-band spectral coverage) and it was
near the point having highest \maxi{} counts. This observation gave us a
wide-band {\bcol data-set} near the peak luminosity with similar energy 
coverage to the \suzaku{} {\bcol data-set} at much lower luminosity levels.
However, as pointed out in section \ref{ss:mod}, the low effective area
of \integral{} {\it JEM-X} resulted in lower signal to noise in the energy band from
11\kev to 17\kev, than for the \suzaku{} data-set. 

\swift{} data reduction was
carried out using \ftools{} from {\tt HEASoft 6.15.1} with {\tt CALDB} updated till 
June 2014. \integral{} analysis was done using {\tt OSA 10.0} with its calibration
files also updated till June 2014. 
We used the \swift{} {\it XRT} data from 0.5\kev to 9.7\kev for our spectral analysis. These data were 
taken in the windowed timing (WT) mode of {\it XRT}. The maximum observed count rate was 
about $40{\rm~counts~s^{-1}}$. This ensured that the {\it XRT} data was not piled up \citep[see][]{rom06}. 
The {\it XRT} data from 0.4\kev to 1\kev did show an excess due to uncertainty in the 
response modelling of {\it XRT} WT mode, but upon using the position dependent
WT response files, this apparent excess was removed\footnote{http://www.swift.ac.uk/analysis/xrt/rmfs.php}.
Hence we used these position dependant response files for all our analysis. 

We used two 
instruments - the {\it JEM-X} and the {\it IBIS} from the {\it INTEGRAL's} suite.
The {\it JEM-X 2} data (from {\it JEM-X}) and the ISGRI data (from {\it IBIS}) were extracted with
higher spectral binning than used in the standard pipeline. The {\it JEM-X} data
were extracted from 3\kev to 30\kev and the {\it IBIS} data from 15\kev to 100\kev,
as per the instructions in the {\it IBIS} Analysis User Manual (Issue 10.0) . The 
latest calibration files were used for generating the spectra and the rebinned rmfs. 
After extraction, the data were grouped and systematic errors 
added as given in Table~\ref{tab:grp}. We used spin phase averaged 
spectra in all the observations. For modelling
the spectra, we used {\tt XSPEC v12.8.1g} \citep{arn96}.

\begin{table}
  \footnotesize
  \centering
  \caption{Grouping scheme and systematic errors used for different spectral datasets.}
  \begin{tabular}{c|c|c}
	\hline
	\textbf{Spectral data} & \textbf{Grouping} & 
	\textbf{Systematic} \\ 
	                       & \textbf{scheme} & \textbf{errors} \\ \hline
	\rxte{} PCU & None & 1\% \\
	\rxte{} {\it HEXTE} & Minimum 60 counts/bin & 1\% \\
	\suzaku{} {\it XIS} & Minimum 60 counts/bin  & None \\
	           & and oversample by 3    &     \\
	\suzaku{} {\it PIN} & Minimum 40 counts/bin & None \\
	\swift{} {\it XRT} & Minimum 60 counts/bin  & None \\
	\integral{} {\it JEM-X}~$^a$ & 31 bins in 3--30\kev  & 3\% \\
	\integral{} {\it IBIS}~$^a$ & 27 bins in 15--100\kev & 1\% \\
	\hline
  \end{tabular}
  $^a$ Coded mask datasets 
	have Gaussian and not Poisson statistics and therefore do not need 
    rebinning.
  \label{tab:grp}
\end{table}

\section{Spectral Analysis}\label{ss:mod}
We began our spectral analysis with the wide-band \suzaku{} dataset (obs-id 406049010) taken
alongwith \rxte{} dataset (obs-id 96032-01-04-00). For other datasets, we used the results
from this analysis as a template because this data-set had both wide-band spectral coverage
and high signal to noise. As noted previously, when
we held the {\it PIN}-{\it XIS} cross-normalisation fixed at $1.16$ we found a significant offset
of only the {\it PIN} spectrum from the other instrument spectrum in the ratio plots
(refer Fig.~\ref{fig:rat}
for the best fit {\it cutoffpl} based model with CRSFs). When we let the normalisation parameter 
free, we found that it gave a best fit value of $1.57$ which we adopted for all our subsequent analysis. 

To model this wide-band spectrum, we started off by using the simple {\it cutoffpl} based continuum
modified by interstellar absorption and CRSFs. At low energies ( $<$ 2 keV), effects of 
local absorption and emission features from the plasma wind of the companion
star are known to affect the spectrum \citep{suc11}. 
However, we did not specifically account for this in the spectrum. 
Complete modelling of the wind spectrum is preferably done with higher spectral resolution
data from gratings. As mentioned in Sec.~2.2, the \suzaku{} data-set 
had calibration uncertainties around the detector's Si K-edge (from $\sim$ 1.6\kev
to $\sim$ 2.5\kev). Thus modelling the wind effects was difficult with
this data-set. As these effects are restricted to lower energies, and as
we do not consider data below 0.8\kev in our analysis, we do not expect
this to change our continuum model estimates significantly. 
The interstellar absorption was modelled by an updated version of \texttt{tbabs}
\footnote{http://pulsar.sternwarte.uni-erlangen.de/wilms/research/tbabs}
using abundances of \citet{wil00} and cross-sections as given in \citet{ver95}.

\begin{figure*}
  \centering
  \includegraphics[height=0.4\textwidth,angle=270]{finexes.ps}
  \includegraphics[height=0.4\textwidth,angle=270]{fincrs3.ps}
  \begin{minipage}{0.4\textwidth}
    \includegraphics[height=\textwidth,angle=270]{fincrs4.eps}
  \end{minipage}
  \begin{minipage}{0.45\textwidth}
    \includegraphics[width=\textwidth]{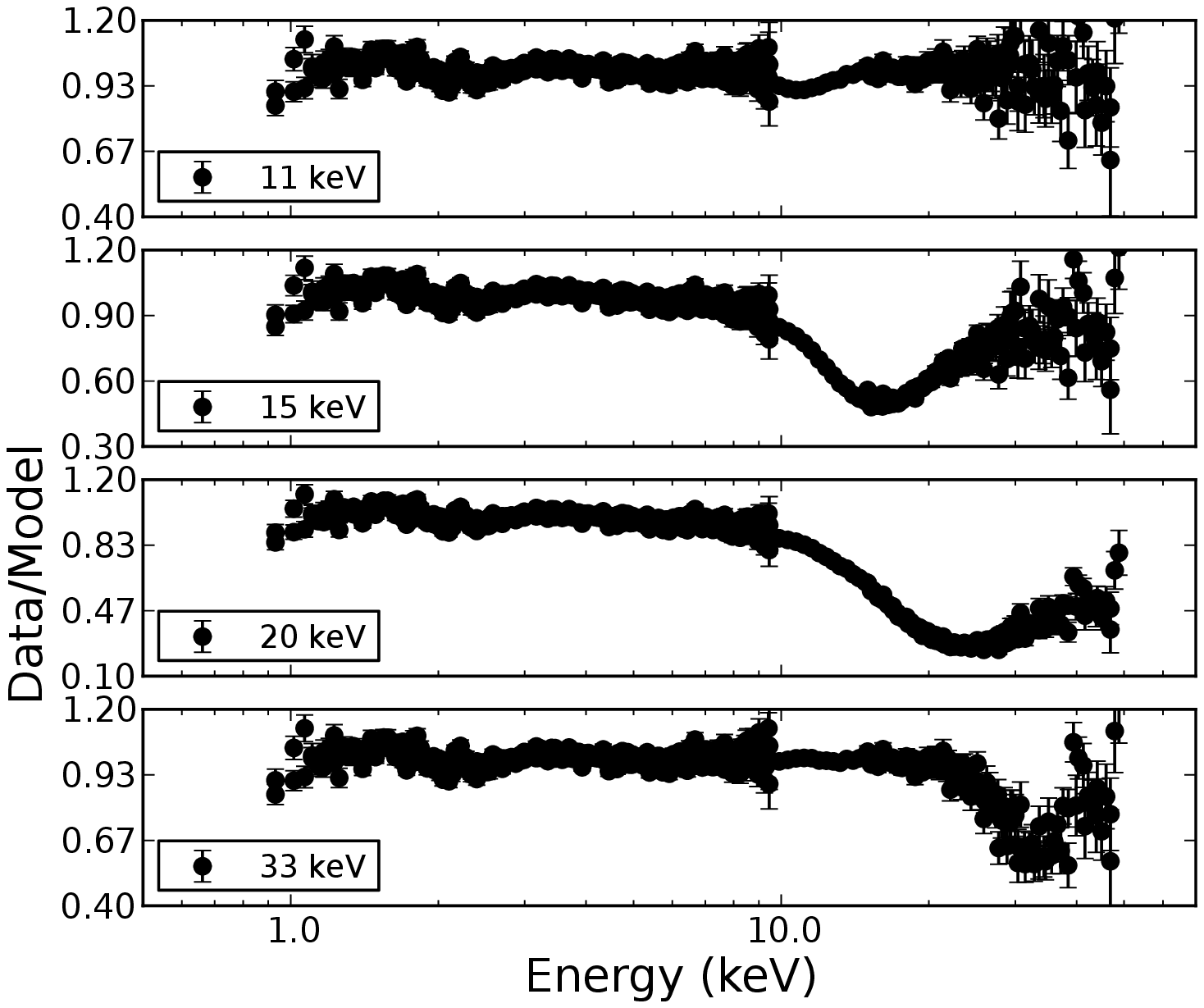}
  \end{minipage}
  \caption{Results of our joint spectral fits to the \suzaku{} {\it XIS} (black
   and red), {\it PIN} (blue) and \rxte{} {\it PCA} (green) and {\it HEXTE} (light blue) 
   data using a cut-off power-law, CRSFs and 
  a narrow iron K$\alpha$ line at $6.4\kev$. Top left: Unfolded spectral data, 
  the model fitted above 3\kev and extended to lower energies in the upper panel and the deviations of 
  the data from the model in the lower panel showing an excess at lower energies. Top right: Similar to top left 
  but the model includes a low temperature blackbody in addition to the cutoffpl and 3 CRSFs.  
  The residuals show possible presence of a 4th cyclotron line at $\sim 11\kev$. Bottom left: Similar to 
  top right but the model includes a 4th cyclotron line at 11 \kev.
  Bottom right: The 4 cyclotron lines as seen in the ratio
   plots obtained by setting the optical depth of the lines to zero after obtaining the best fit. See online 
   edition for a color print of this figure.}
  \label{fig:exs}
\end{figure*}

This model fit well if we took data above 3\kev, but showed a significant 
excess when we included the data from 0.8\kev to 3\kev. This is clearly
seen in the top left panel of Figure \ref{fig:exs}, where we obtained $\chi^2/dof = 2018.60/474$.
On using a blackbody {\tt (bbody)} to account for this excess, we found that the spectral data were 
well fitted to give a significant improvement in fit $\chi^2/dof = 731/472$.
The temperature and radius of the blackbody so obtained are listed in Table \ref{tab:bb}.
The radius was computed from the normalisation of the blackbody  model {\tt bbody}, 
which depends on 
the luminosity and distance to source. The distance was taken as ${7\rm~kpc}$ and luminosity
was taken as ${\rm L_{bb} = \sigma .\, T^4 .\, Area}$.
The blackbody radius is clearly a factor of 10 larger than expected values for a hotspot 
at the base of the accreting mound. However, given that we might be integrating 
emission area from the base of the column and that effects of
comptonization of
this blackbody due to the accreting plasma are not considered, it is likely that
these values are over-estimates. We did try the {\tt XSPEC} model given by \citet{far12}
to see if this was true and found values of blackbody radius to drop to about
6 kms (see Appendix \ref{ap:bb} for details). The large number of free parameters in this model, though, make it
difficult to use while testing the CRSF parameters across multiple
data-sets. 

We also tried
using {\tt comptt} model to account for this low energy excess, as indeed
was tried in \citet{fer09}. This gave us an improvement in fit with the
$\chi^2$ reducing from 688/469 to 583/466. This large improvement strongly
indicates that the blackbody seed photons are indeed comptonized. The spectral
fits for {\tt comptt} were obtained though, for a electron plasma temperature (kT${\rm_e}$) of 1.2 \kev, a seed
photon temperature of 0.2 \kev and plasma optical depth($\tau$) of 35. The
{\tt comptt} code itself is meant for use for plasma temperature greater than
2 \kev and at such high $\tau$, the shape of the spectrum is very similar to
the Wien tail of a blackbody. Using the {\tt comptt} model for this Wien tail alone,
also lead to an increase in the value of the absorption 
column from $N_H=1.3\times 10^{22}{\rm~cm^{-2}}$ to $N_H=1.7\times 10^{22}{\rm~cm^{-2}}$.
When we tried the same model in the {\it Swift /
INTEGRAL} spectrum, we found a very small reduction in the fit statistic $\chi^2$
from 146/166 to 145/163. Furthermore, none of the {\it RXTE} observations gave an improvement 
in fit $\chi^2$ when using the {\tt comptt} model as compared to the {\tt bbody} model. 
Finally, we found very little differences in the CRSF energies, while
using the simple blackbody model as compared to the comptonized blackbody. Since the 
data that we analysed could not help us pin down the nature of the low energy continuum, we
used the simple blackbody for all subsequent analysis. However, we do note that
probing this component further may be important to understand the overall continuum, which 
in turn may affect the cyclotron line results.

As stated in \citet{suc11} and \citet{mul13}, we found the column density
($N_H$) to be strongly correlated with the power-law index ($\Gamma$)
and to be varying across observations. To prevent any unwanted effects due 
to changing $N_H$ on our fits, we fixed the column density for all
observations to the best fit value of $N_H=1.3\times 10^{22}{\rm~cm^{-2}}$
obtained from the wide-band \suzaku{} observation.
We also used a narrow emission line with its 
centroid fixed at 6.4\kev and width fixed at 10$^{-4}$\kev to model the 
Fe K-$\alpha$ fluorescence emission which has been observed in many other 
HMXB NS binary systems and is also expected to be seen in 4U~0115+63
\citep{tor10,mul13}. For the wide-band \suzaku{} observation, this narrow
Fe-line component 
gave an improvement in $\chi^2/dof$ from 756/473 to 731/472. Upon testing
this with the {\it simftest} script of {\tt XSPEC}, we found it to be
significant to a level greater than 3$\sigma$ for 480 iterations. This test was done using XIS data only, to 
isolate any effects of the continuum beyond 10 \kev.

\begin{table}
  \centering
  \caption{Blackbody radius and area for different observations. Note that \rxte{}
  observations have higher errors.}
  \begin{minipage}{0.5\textwidth}
  \begin{tabular}{lccc}
	\hline
	{Instrument} & {Day (MJD)} &  kT (\kev)\footnote{All errors in this table are 1$\sigma$ deviations} & {Radius (kms)}\footnote{calculated as $\sqrt{\frac{area}{\pi}}$} \\
	\hline
	\rxte{} & 55730.06 & $1.05^{-0.18}_{+0.19}$ & $13.12^{-3.78}_{+5.66}$ \\ 
	\swift{} / \integral{} & 55736.34 & $0.93^{-0.19}_{+0.41}$ & $17.39^{-4.31}_{+9.22}$ \\
	\rxte{} & 55738.87 & $0.94^{-0.29}_{+0.65}$ & $17.38^{-6.65}_{+15.36}$ \\ 
	\rxte{} & 55739.51 & $1.09^{-0.0}_{+0.0}$ & $14.69^{-0.0}_{+0.0}$ \footnote{\label{fn1}Could not put error constraints on blackbody temperature} \\ 
	\rxte{} & 55743.91 & $0.79^{-0.20}_{+0.21}$ & $19.57^{-6.24}_{+6.47}$ \\ 
	\rxte{} & 55747.90$^{\ref{fn1}}$ & $0.78^{-0.0}_{+0.0}$ & $18.85^{-0.0}_{+0.0}$ \\ 
	\suzaku{} / \rxte{} & 55750.82 & $0.73^{-0.01}_{+0.01}$ & $13.95^{-0.42}_{+0.41}$ \\ 
	\rxte{} & 55751.32 & $0.92^{-0.26}_{+0.38}$ & $10.54^{-3.61}_{+5.37}$ \\ 
	\hline
  \end{tabular}
  \end{minipage}
  \label{tab:bb}
\end{table}

We model the CRSFs using a symmetric Lorentzian like absorption feature 
{\tt cyclabs} \citep{mih90}, with each line energy being independently
determined by the fit. This was done to account for anharmonic line 
ratios which can occur due to effects of viewing 
geometry and asymmetric emission patterns, as has been reported in 
other HMXB sources \citep{kre02,pot05} and in theoretical simulations \citep{nis05,sch07,muk12,muk13b}.

\begin{table*}
  \centering
  \caption{Best-fit parameters for the CRSFs.}
  \begin{minipage}{\textwidth}
  \begin{tabular}{cccccccc}
	\hline
	{\bf Day (MJD)} & {\bf Ecyc (\kev)} & {\bf FWHM (\kev)} & {\bf $\tau$} &
	{\bf $\chi^2$/dof } & {\bf Eq. width (\kev)} & {\bf $\Delta\chi^2$ /
	$\Delta$dof}\footnote{Change / reduction in $\chi^2$ on addition of extra CRSF / change in number of dof
	 - denotes the confidence that a 4 CRSF model is better than a 3 CRSF model}
	 & \bf F-stat \%\\
	\hline
	55730.06     & $16.33^{-0.32}_{+0.59}$ & $9.38^{-0.76}_{+0.56}$ & $1.43^{-0.13}_{+0.16}$ & 57 / 73 & 11.98 $\pm$ 0.72  & 0 / 0 & 0\\
	(\rxte{})	 & $37.29^{-1.15}_{+1.08}$ & $4.0^{-0}_{+0.0}$ & $0.45^{-0.14}_{+0.14}$ &           & 2.38\footnote{Cannot quote error on this, as width is frozen \label{ft:tab}} & & \\
	55736.34     & $10.31^{-4.58}_{+1.05}$ & $5.96^{-3.63}_{+4.58}$ & $0.97^{-0.62}_{+0.31}$ & 146 / 166 & 6.21 $\pm$ 4.23  & 0 / 0 & 0\\
	(\swift{} / \integral{}) & $20.00^{-8.21}_{+6.07}$ & $9.92^{-8.97}_{+5.24}$ & $1.26^{-0.90}_{+0.31}$ &           & 11.98 $\pm$ 4.50 & & \\
				 & $35.75^{-2.19}_{+4.17}$ & $4.0^{-0}_{+0.0}$ & $0.45^{-0.36}_{+0.38}$ &           & 2.38$^{\ref{ft:tab}}$ & & \\
	55738.87     & $11.71^{-0.49}_{+0.37}$ & $4.33^{-1.47}_{+1.22}$ & $0.62^{-0.3}_{+0.25}$ & 47 / 67  & 3.35 $\pm$ 1.18 & 14.91 / 3 & 83\\
	(\rxte{})    & $15.82^{-0.68}_{+0.57}$ & $1.1^{-1.06}_{+2.97}$ & $0.1^{-0.06}_{+0.31}$ &  & 0.17 $\pm$ 0.19 & & \\
				 & $19.86^{-0.68}_{+1.01}$ & $8.82^{-1.62}_{+1.91}$ & $1.21^{-0.16}_{+0.38}$ &  & 10.48 $\pm$ 1.59 & & \\
				 & $35.53^{-1.65}_{+1.78}$ & $4.0^{-0}_{+0.0}$ & $0.35^{-0.14}_{+0.16}$ &  & 1.93$^{\ref{ft:tab}}$ & & \\
 	55739.51     & $11.66^{-1.09}_{+0.39}$ & $5.12^{-3.7}_{+1.98}$ & $0.74^{-0.44}_{+0.2}$ & 43 / 67 & 4.49 $\pm$ 4.56 & 5.81 / 3 & 64\\
	(\rxte{})    & $16.08^{-2.16}_{+1.01}$ & $2.09^{-2.05}_{+4.53}$ & $0.14^{-0.1}_{+0.98}$ &  & 0.42 $\pm$ 1.40 & & \\
				 & $20.42^{-2.16}_{+2.17}$ & $8.52^{-2.26}_{+1.25}$ & $1.31^{-0.22}_{+0.13}$ &  & 10.58 $\pm$ 3.59 & & \\
				 & $35.24^{-1.38}_{+1.51}$ & $4.0^{-0}_{+0.0}$ & $0.38^{-0.14}_{+0.15}$ &  & 2.07$^{\ref{ft:tab}}$ & & \\
	55743.91     & $11.25^{-0.98}_{+1.03}$ & $4.13^{-2.37}_{+5.04}$ & $0.45^{-0.33}_{+0.32}$ & 31 / 67  & 2.50 $\pm$ 2.94 & 10.17 / 3 & 84\\
	(\rxte{})    & $15.2^{-1.29}_{+0.72}$ & $2.82^{-2.76}_{+3.08}$ & $0.38^{-0.27}_{+0.57}$ &  & 1.44 $\pm$ 2.35 & & \\
				 & $19.56^{-1.29}_{+1.97}$ & $7.57^{-1.73}_{+1.39}$ & $1.17^{-0.42}_{+0.11}$ &  & 8.86 $\pm$ 2.09 & & \\
				 & $32.77^{-0.85}_{+0.94}$ & $4.0^{-0}_{+0.0}$ & $0.6^{-0.14}_{+0.15}$ &  & 2.97$^{\ref{ft:tab}}$ & & \\
	55747.90	 & $11.20^{-0.86}_{+1.04}$ & $3.53^{-2.01}_{+4.45}$ & $0.33^{-0.19}_{+0.39}$ & 52 / 67 & 1.63 $\pm$ 1.11 & 14.99 / 3 & 81\\
	(\rxte{})             & $14.93^{-0.98}_{+0.83}$ & $2.76^{-2.1}_{+2.41}$ & $0.35^{-0.28}_{+0.51}$ &  & 1.33 $\pm$ 1.40 & & \\
				 & $18.01^{-0.98}_{+2.54}$ & $9.42^{-1.72}_{+1.67}$ & $1.33^{-0.32}_{+0.1}$ &  & 11.67 $\pm$ 1.34 & & \\
				 & $33.79^{-1.96}_{+2.52}$ & $4.0^{-0}_{+0.0}$ & $0.26^{-0.14}_{+0.15}$ &  & 1.46$^{\ref{ft:tab}}$ & & \\
	55750.82     & $10.77^{-0.27}_{+0.32}$ & $0.96^{-0.56}_{+0.67}$ & $0.12^{-0.03}_{+0.04}$ & 688 / 469  & 0.17  $\pm$ 0.10 & 42.68 / 3 & 74\\
	(\rxte{} / \suzaku{}) & $14.39^{-0.33}_{+0.35}$ & $4.19^{-0.92}_{+0.97}$ & $0.69^{-0.25}_{+0.22}$ &  & 3.48 $\pm$ 1.60 & & \\
				 & $19.53^{-0.33}_{+1.48}$ & $8.33^{-1.52}_{+1.56}$ & $1.13^{-0.19}_{+0.14}$ &  & 9.52 $\pm$ 2.21 & & \\
				 & $31.35^{-0.84}_{+0.98}$ & $4.0^{-0}_{+0.0}$ & $0.56^{-0.16}_{+0.16}$ &  & 2.83$^{\ref{ft:tab}}$ & & \\
	55751.32 	 & $10.76^{-0}_{+0.0}$ & $2.1^{-2.1}_{+7.33}$ & $0.12^{-0.09}_{+0.59}$ & 57 / 67 & 0.37  $\pm$ 0.59 & 4.32 / 3 & 55\\
	(\rxte{})    & $14.51^{-0.84}_{+0.95}$ & $4.67^{-2.91}_{+3.08}$ & $0.73^{-0.28}_{+0.61}$ &  & 4.06 $\pm$ 1.07 & & \\
				 & $19.55^{-0.84}_{+0.0}$ & $8.7^{-1.62}_{+13.52}$ & $1.34^{-0.62}_{+0.25}$ &  & 10.90 $\pm$ 1.84 & & \\
				 &  $31.98^{-1.16}_{+1.56}$ & $4.0^{-0}_{+0.0}$ & $1.05^{-0.7}_{+0.42}$ &  & 4.44$^{\ref{ft:tab}}$ & & \\
	\hline
  \end{tabular}
\end{minipage}
  \label{tab:lin}
\end{table*}

\begin{table}
  \centering
  \caption{Continuum parameters for the two wide-band observations}
  \begin{tabular}{ccc}
	\hline
	{\bf Parameter} & {\bf \swift{} / \integral{}} & {\bf \suzaku{} / \rxte{}} \\
	\hline
	\nh ($10^{22}$ cm$^{-2}$) & 1.3 & 1.3 \\
	$\Gamma$ & $-2.16^{-1.10}_{+0.48}$ & $-1.25^{-0.08}_{+0.07}$\\
	$E_{cut}$\kev & $4.81^{-0.99}_{+1.22}$ & $5.34^{-0.20}_{+0.21}$\\
	$kT_{bb}$\kev  & $0.94^{-0.19}_{+0.41}$ & $0.73^{-0.01}_{+0.01}$ \\
	$bb_{norm}$ (x $10^{-3}$) & $15.5^{-1.7}_{+1.6}$ & $3.7^{-0.2}_{+0.2}$ \\
	Fe line eq.width (\ev) & $11.7^{-11.7}_{+20.5}$& $11.7^{-3.0}_{+3.4}$\\
    \hline
  \end{tabular}
  \label{tab:sum}
\end{table}

Modelling the \suzaku{} spectrum, we found cyclotron lines to be 
at $\sim$ 14 \kev, $\sim$ 21 \kev and $\sim$ 31 \kev. However, we additionally
noted an absorption feature at 11\kev in the fit residuals. Modelling
this using a Lorentzian like absorption feature gave a
significant improvement in the fit as seen in Fig. \ref{fig:exs}, with the
$\chi^2/dof$ changing from 731/472 to 688/469.
We were not able to fit for any CRSF features above 35\kev in any of the datasets
because of large noise at the higher energies. And the CRSF at $\sim$ 33\kev 
was itself constrained by keeping its FWHM fixed at 4\kev \citep{fer09,mul13}. The 4 CRSFs
obtained in this observation (as listed in Table~\ref{tab:lin}) are at about 
11\kev, 15\kev, 20\kev and 33\kev.

We tried next to model the high luminosity wide-band data of \swift{} and \integral{} 
simultaneous observation. We were able to fit the spectrum from 0.5\kev upto 
60\kev using only the {\it cutoffpl} continuum with absorption fixed at
$N_H = 1.3 \times 10^{22}{\rm~cm^{-2}}$ and CRSFs ( $\sim$ 11\kev,
20\kev and 33\kev) to get a $\chi^2/dof = 172.8/168$. 
On adding the blackbody component, this improved to  $\chi^2/dof = 146.1/166$. 
We tested the significance of this using the {\it lrt} script of {\tt XSPEC}. 
This gave a significance of greater than 3$\sigma$ for the blackbody model for 408 iteration runs. 
When taken alongwith the case of {\it Suzaku / RXTE} data-set, we see that
the presence of a soft X-ray component is justified. As stated above, modeling this 
component is important. For the sake of consistency, we use the blackbody model, as using the
{\tt comptt} model for the soft X-ray component did not give us any improvement over the blackbody based
model for this observation.
While using this blackbody and cutoffpl continuum, we could not fit this data-set with the 5th line as earlier
and the fit was consistent with only 3 CRSFs. A summary of the continuum 
parameters obtained for the two wide-band data-sets is given in Table~\ref{tab:sum}.
Errors, unless other-wise mentioned are quoted for a level of 90\% 
confidence in all cases. 

We finally analysed the set of standalone \rxte{} observations.
Due to the lack of coverage of the low energy bands in these data-sets, it was
difficult to constrain the blackbody parameters. Hence we fit the blackbody by
starting from a guess value based on the fit parameters obtained from the
wide-band data-set and let the fit routine converge to give the
values as quoted in Table~\ref{tab:bb}. The blackbody temperature was
frozen to this best fit value while computing errors on other fit 
parameters. Excepting one, all the \rxte{} {\bcol data-sets}
gave an improvement in the fit statistic when using 4 CRSFs as
compared to the fit with 3 CRSFs (see Table~\ref{tab:lin}). The 
exception was the first data-set, taken during the rising phase, which
needed only two CRSFs to describe the spectrum. We consider this
case separately in the Sec.~4. Given this improvement when using 4 CRSFs,
we decided to test its statistical significance for describing the
spectral data.

\subsection{Statistical significance of the 4$^{th}$ cyclotron line}

Computing detection significance of CRSFs must be handled differently from 
the standard techniques usually employed for emission features.  First, following 
the method outlined in \citet{orl12} we construct the F-statistic directly 
as a ratio of the normalised fit $\chi^2$ \citep{orl12,bev92}, as against using 
the standard F-test implemented in {\tt XSPEC}, which constructs the F-stat as a 
ratio of change in normalised $\chi^2$ to the original $\chi^2$.

We try to compute the significance of the 4th CRSF at 11 \kev, which
we used as stated previously to improve the fit.
For the \rxte{}/\suzaku{} joint spectral fit, the above method gave us an F-statistic 
value of $1.06$ with a probability value for
occurrence due to random noise (PCI) equal to $0.26$. We also noted that ignoring
data from 1.6 to 2.5\kev (the energy range with calibration mis-match between
the two {\it XIS} detectors) reduces this probability to 0.24 with very little change 
in spectral parameters. This is not enough to claim detection of a 4th 
cyclotron line. The F-stat based test performed above, looks at what fraction of the
variance of the data-set can be explained by the model. This test looks at the
percentage of data variance that a new model (with all its components) can
explain as compared to the percentage of data variance explained by the old
model (with all its components). Unaccounted residuals exist in the
soft part of our fit spectrum below 3\kev{}. This might account for the low
reduction in percentage variance accounted for by the addition of a 
cyclotron line at
the higher energies (11 \kev) . We checked each of the other \rxte{} observations
individually with this F-test (results listed in Table
\ref{tab:lin}). 
The broadband continuum is not completely sampled by the \rxte{} data alone
and this could explain the low improvement in $\chi^2$ for the 
individual \rxte{} data-sets. To improve the statistics,
we did a joint fitting of all the data-sets using both the 3 CRSF and the 4 CRSF
models. In doing this fit, we let the line energies for each observation be independently
estimated by leaving them untied. This gave an F-statistic value of 1.07 and a
probability of chance occurrence of 0.15 (or a confidence level of 85 \%). 

Finally, we tried a more robust numerical evaluation of significance 
from Monte-Carlo simulations. We did this by finding the probability of false detection of 
the 4th CRSF, assuming the 3 CRSF model to be true. By simulating spectra 
using the {\tt XSPEC} {\it lrt} {\bcol script}, we generated a large
number (7438) of simulated data-sets following the
continuum model with 3 CRSFs modified by statistical noise. 
The {\it lrt} {\bcol script} generates
these data-sets by using the fit covariance matrix to make a random draw of the 
fit parameters. The model so obtained, is convolved with the response matrices of the
individual detectors, and statistical (Poisson) noise is added to each such
simulated {\bcol data-set}. We searched for the presence of the 4th cyclotron line in each of these 
data-sets by trying to fit them with the continuum and 4 cyclotron lines, and compared it
to a fit with the continuum and 3 cyclotron lines.
We tabulated the $\chi^2$ fit value for each such effort. To check if the 
observed 4th CRSF was significant, we compare the {\bcol statistic of each of the simulation
runs against the statistic of the observed data}. The form of the {\bcol statistic we used 
was the F-stat ($fst$) constructed as mentioned before \citep[see also][]{sar15}}. We obtained three instances where 
the simulated $fst$ was as high as the observed $fst$. However, in none of these instances, 
the 4th CRSF fit with centroid near 11 \kev. This gave a PCI of 3 in 7438, or a significance of 
3.5$\sigma$ for the observed CRSF at 11 \kev.

This result, though needs to be treated with care because of the poor 
$\chi^2$ of the original fit (731/472) v/s (688/469). We proceed assuming the 
4 CRSF model to be the better one, but only after noting that the poor initial fit 
could influence the Monte-Carlo results.

\begin{figure}
  \centering
  \includegraphics[width=0.5\textwidth]{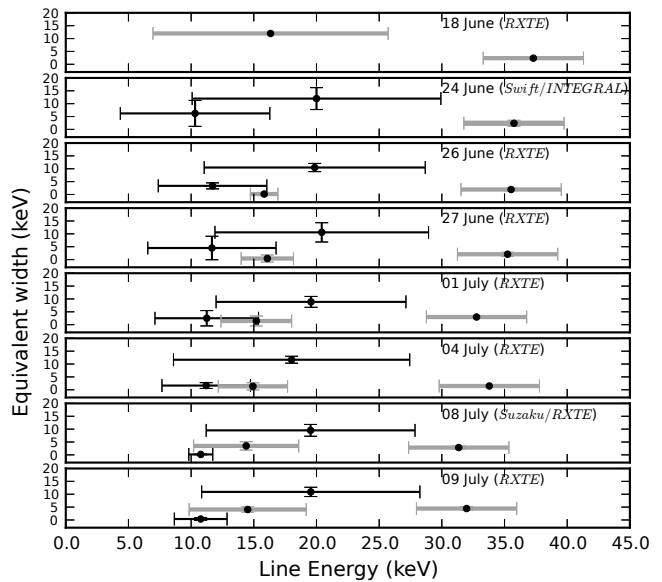}
  \caption{Line Distribution diagram. This figure shows the variation in 
	energy, FWHM and equivalent widths of all the lines we obtained in 
    each of our observations. The central point represents the line energy
	and equivalent width with horizontal bars depicting the FWHM and vertical
    bars representing errors on the equivalent width. We use black to indicate
  the 11\kev line set and gray to indicate the 15\kev line set.}
  \label{fig:dis}
\end{figure}

We then looked at the validity of the 4 CRSF model in describing other data-sets.
All the individual \rxte{}
spectra do not have wide-band coverage and have lower spectral resolution than
say the combined \suzaku{} / \rxte{} data-set. This could be a reason
for the marginal improvement in the $\chi^2$ statistic.
The fact that multiple observations showed
signs of 4 CRSF features, with an improvement in $\chi^2$ would lead to an 
increased relevance of this detection. Additionally, given that each of these observations gave
nearly similar centroid energies for all 4 CRSFs (see Fig.~\ref{fig:dis}) would also
increase the significance of our detection.

The possible reasons for 
two observations not requiring 4 CRSFs to describe their spectra are 
discussed herewith. The \swift{} and \integral{} simultaneous observations
not fitting with the 
4 CRSF model could be either because the line at 15\kev does not exist in this
observation, or because the line is too weak to be detected by the {\it JEM-X} detector
due to its much lower collection area and poorer signal to noise than the
  \rxte{}/{\it PCA}.
This is where we find the high signal to noise of the combined \suzaku{} / \rxte{} data
to be critically important. The \rxte{} standalone observation which did
not show 4 CRSFs, though had a high signal to noise all through the 
expected CRSF energy ranges. We take the detection of only 2 CRSFs in this
observation to be a valid result and discuss the possible reasons for non detection in 
this data-set in the next section.

A useful indicator for evaluating the strength of the absorption lines, and
to see if they are physically relevant is the line equivalent width. We computed the 
equivalent width as 

\begin{eqnarray}
  {\rm EW \; (keV)} = \int_{E1}^{E2} (1 - e^{(-\tau P(E))})\, {\rm d}E \\ 
  P(E) = \frac{(W\frac{E}{E_{cyc}})^2}{(E - E_{cyc})^2 + W^2} \label{eqn:lew}
\end{eqnarray}

where $P(E)$ is the line profile that we use, $W$ is FWHM of the line, 
$E_{cyc}$ is its energy and $\tau$ is its depth. By simple error propagation 
of the variances obtained from the fit covariance matrix, we can get an idea 
of errors on the estimate. We calculated this for all 4 CRSFs to check 
for their relevance. The results of {\bcol such} calculations are plotted in Fig.~\ref{fig:dis}. 
We try to make sense of these results in the next section.

\begin{figure}
  \centering
  \includegraphics[width=0.5\textwidth]{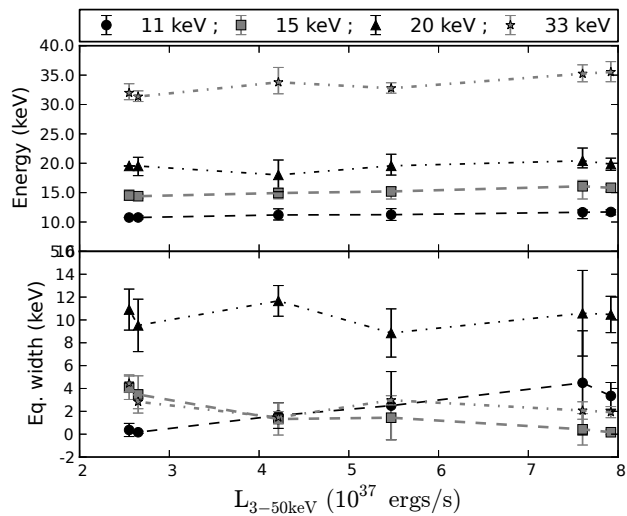}
  \caption{4 CRSFs and their trends with changing luminosity. Top panel shows variation in CRSF
  energy with luminosity for each of the lines in 4 CRSF model. Bottom panel shows
  variation in equivalent width for the same set of lines. These plots are made for
  all data-sets which can be described using 4 CRSFs.
  }
  \label{fig:tbd}
\end{figure}

\begin{figure}
  \centering
  \includegraphics[width=0.5\textwidth]{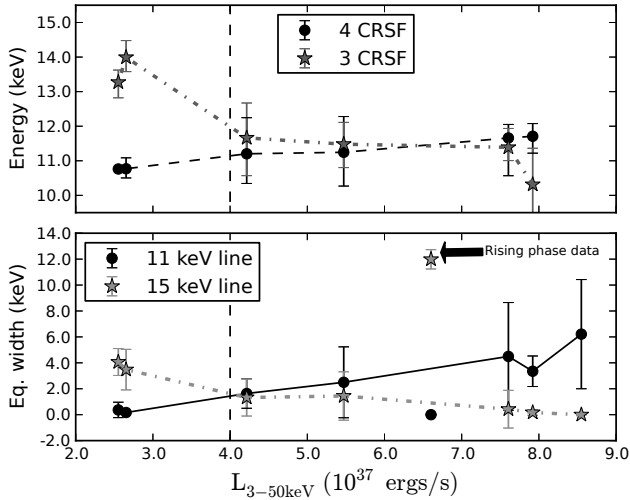}
  \caption{Trends in the fundamental lines.Top panel shows the effect of not using 4 CRSF
  lines, which leads to the anti-correlation with luminosity as reported in previous works.
  Bottom panel shows the change in line equivalent width of the fundamental of each set of the 
  4 CRSF model. 
  The vertical line is placed at the same luminosity level as in Fig~\ref{fig:obs}.
  The rising phase {\bcol data-set}, which does not seem to follow the trend is marked separately.
  See text for details.}
  \label{fig:tre}
\end{figure}

\section{Discussion}\label{ss:con}
\subsection{Summary of results}
In this work we have analysed the spectra from different X-ray observatories 
covering the 2011 outburst of the HMXB 4U~0115+63. These include eight
observations spread across twenty days including two wide-band observations
comprising of data from multiple satellites (\swift{} + \integral{} and
\suzaku{} + \rxte{}). The use of multiple instruments covering different 
energy ranges enabled us to properly model the broadband continuum, which
in turn yielded better constraints on the CRSFs. The two major results from our analysis are:

\begin{enumerate}
  \item {\bf Broadband modelling of the continuum:} The continuum was well constrained 
	with a blackbody component 	modelling the soft X-ray bands and a cut-off power-law for 
	the harder bands. The blackbody component was found to be statistically significant 
	in both of the wide-band observations, as well as the standalone \rxte{} observations. 
	Excess emission features in the shape of a Gaussian, as reported in previous
	works  e.g. \citet{mul13} was not required for our analysis. Such Gaussian features have previously 
	been attributed to possible cyclotron emission \citep{bec07,fer14}, however such results 
	were inconclusive.

  \item {\bf Two sets of cyclotron lines:} From the spectral analysis we find evidence for presence 
	of two independent sets of cyclotron lines, each with a harmonic, with fundamental energies 
	centred at  $\sim 11$ \kev line and  $\sim 15$ \kev line. Previous works find either one of 
	the two fundamental lines, with the 15 \kev line found at lower luminosities, and the 11 \kev 
	line found at higher luminosities \citep{tsy07}, resulting in the inference of 
	anti-correlation of the line energy and luminosity. To check whether the 4 cyclotron lines are 
	two independent sets of harmonics we compared the variation of equivalent widths of the lines 
	evaluated following eq.~\ref{eqn:lew}. We also investigated variations in line energies with
	the outburst luminosity (summarised in Figures~\ref{fig:dis}, \ref{fig:tbd} and \ref{fig:tre}). 
	We took the line equivalent width of the non-detected line to be zero. From the figures we note that: 
	\begin{enumerate}
	  \item There is no luminosity dependence of the fundamental 
		cyclotron line at 11\kev if we use the
		 4 CRSF model, where both the $\sim 11$ and  $\sim 15$ keV lines are present. The anti-correlation
		 appears if the 3 CRSF model is used, as shown in the top
		 panel of Fig. \ref{fig:tre}. This could be an artifact  resulting from incorrect spectral
		 modelling with only 3 cyclotron lines, instead of 4 CRSFs, as we discuss below.
	  \item In the 4 CRSF model, the line equivalent widths of the two lines
		at 15\kev and 11\kev show opposite variations with luminosity as seen
		in the bottom panel of Figure \ref{fig:tre}.
	\end{enumerate}

When only 3 cyclotron lines are used to model the spectra, the two lines at 
$\sim$ 11\kev and $\sim$ 15\kev are modelled by a single CRSF component. Since 
the equivalent widths of these two lines changes
with luminosity, the single averaged CRSF component is closer to the
line with higher equivalent width in the given observation.
Simulations have shown that often the second harmonic is deeper and more prominent \citep{ara00,sch07}
due to photon filling and emission by de-excitation near the fundamental 
CRSF energy. Results from previous outbursts \citep{li12,bol13}
and our analysis demonstrate the near constant line energy of the 20\kev line.
This result too would lead us to expect a near constant line energy for the
fundamental 11\kev line.

If we split the 4 CRSFs that we obtained into two sets of 
harmonics with one at (11\kev and 20\kev) and the other at (15\kev and 33\kev), 
it helps us explain our observations as listed below : 

	\begin{enumerate}
	  \item As shown, this explains the reason for observations of an anti-correlated
		fundamental CRSF with luminosity.
	  \item It could be the reason why 
	this source is the only one to have shown multiple (upto 5) harmonics of the
	fundamental cyclotron line 
	in its spectrum. If we have two such line forming regions with fundamentals 
	at 11\kev and 15\kev, and each region showed the presence of 2 harmonics, 
	then they can easily be confused for multiple harmonics from a single 11\kev
	fundamental CRSF. 
	  \item This would explain why the CRSFs we obtain at 
	20\kev varies so litle, whereas the one we get at 33\kev varies a lot more.
	Under our hypothesis, the $\sim$ 33\kev line in our observations would be a combination of
	the 2nd harmonic from the 15\kev set (at $\sim$ 30\kev) and the 3rd harmonic from the 
	11\kev set (at $\sim$ 33\kev ). As seen from 
	Fig.~\ref{fig:dis} and \ref{fig:tbd}, the $\sim$ 33\kev line shifts to higher energies
	when the $\sim$ 11\kev line becomes stronger. 
	  \item Finally, it gives a plausible reason for detection of only two CRSFs in the \rxte{} 
	observation taken during the rising phase. If, for some reason, the 11\kev line set 
	is either not present or has very weak signatures in the rising phase of the outburst, 
	then the observed CRSF energies in the rising phase would correspond to the 
	15\kev line set only. The values we get for the observed CRSFs seem 
	very close to this.
	\end{enumerate}

\end{enumerate}

\subsection{Possible origin of the cyclotron line sets}
In this section we discuss where the two scattering regions could
be located and what causes the CRSF strengths from these regions
to vary in such a manner. 

The first possibility is emission from two 
different regions at different heights on the
same pole. If we assume a dipole like magnetic field
structure of the Neutron Star (NS) the difference of $\sim$ 4\kev between
two sets of lines can be caused if the difference in height
of the emitting regions is $\sim$ 1.1 km, which is approximately the shock height above the
neutron star surface \citep{bec12}. If the 15\kev line set originates in the fan beam at
the base of the mound, and the 11\kev line set in the pencil beam from the
top of the shock, then variation in line equivalent widths with luminosity can 
be explained in terms of varying strengths of the fan and pencil beam emissions. 
At higher luminosities, the fan beam is expected to dominate, whereas both beams will 
be visible at intermediate luminosities
levels \citep[see Fig.~1 of ][]{bec12}. However, the effect of such variations of luminosity 
on the equivalent width is unclear and not adequately addressed by existing theoretical models.

\begin{figure*}
  \includegraphics[width=6cm]{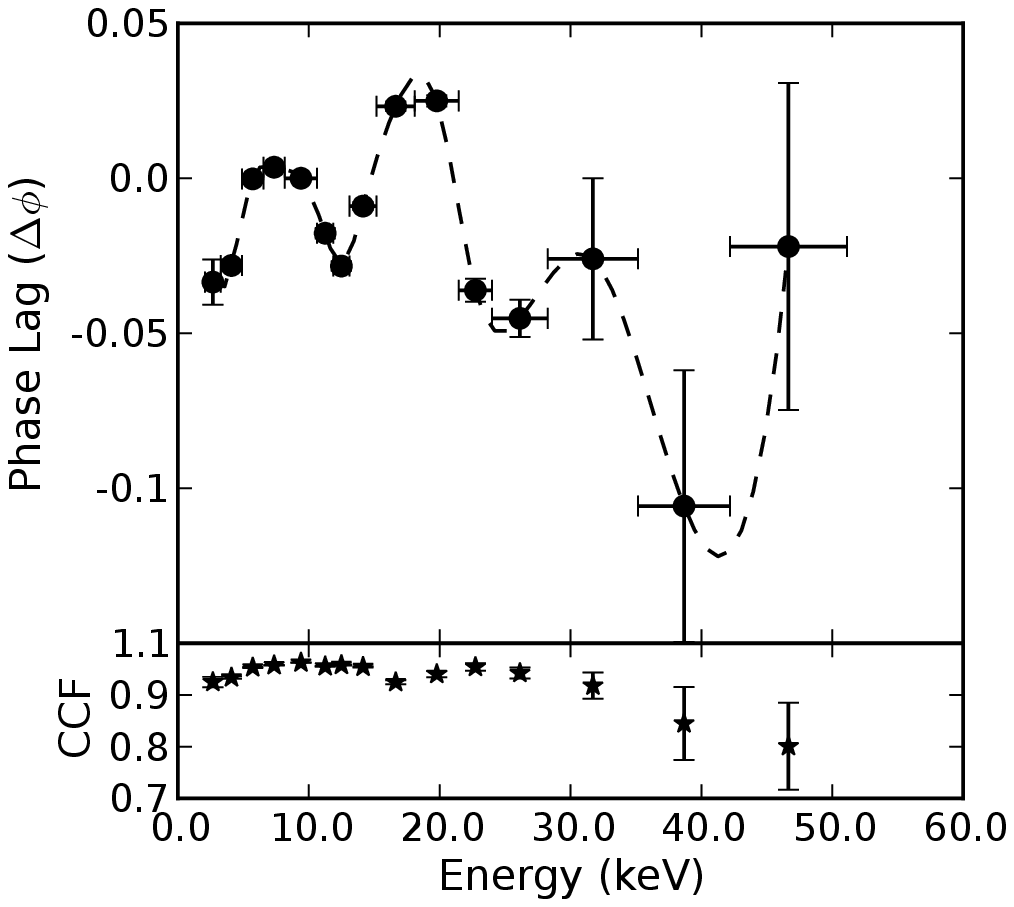}
  \qquad
  \includegraphics[width=6cm]{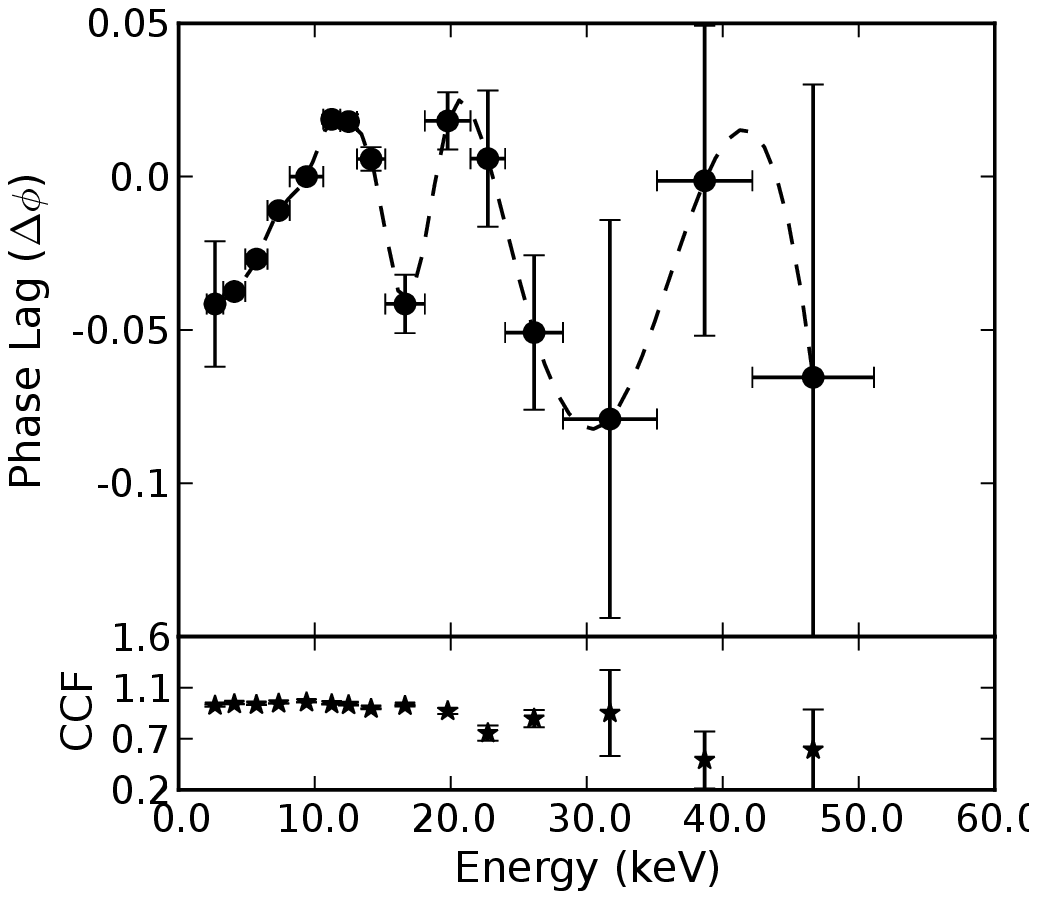}
  \quad \\
  \includegraphics[width=6cm]{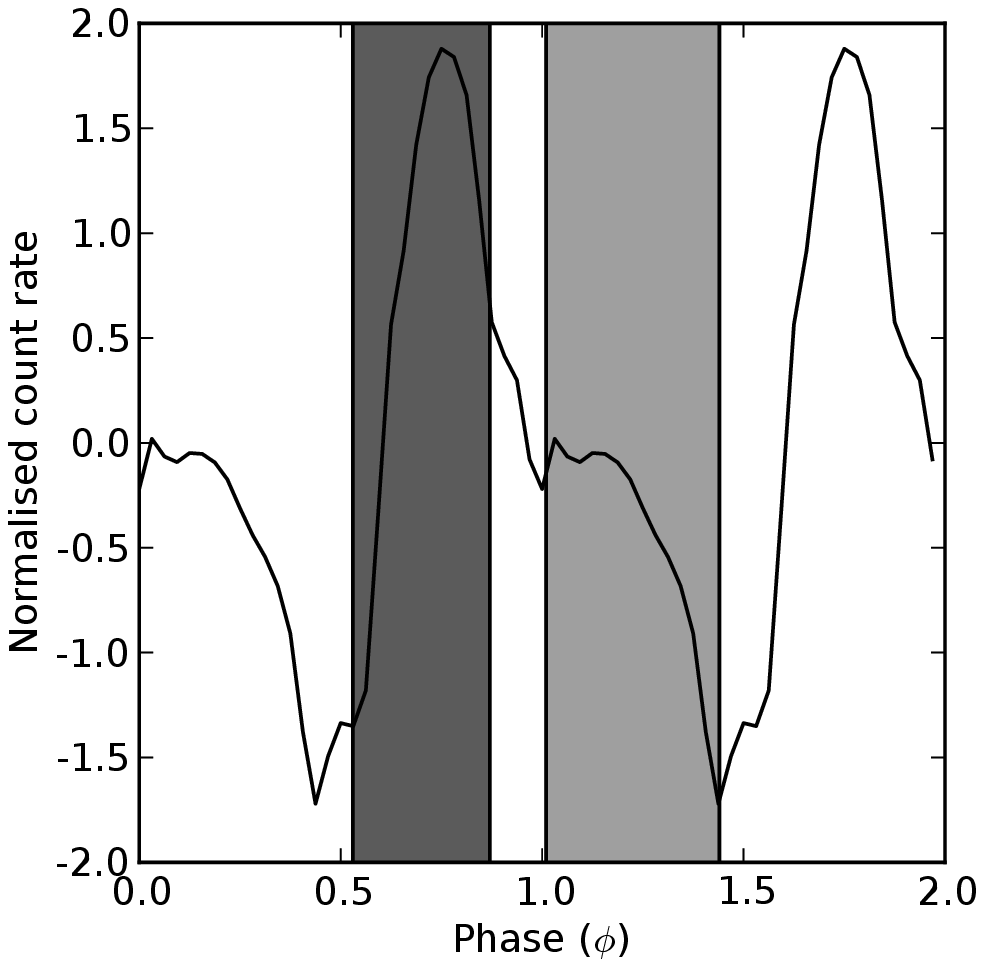}
    \caption{Phase lag spectrum and the pulse profile. Bottom panel shows the 
	pulse profile with shaded regions depicting the phase-bins over which correlation
    was computed. Top left panel is the phase-lag spectrum (and the correlation coefficient)
    for the Main peak (shaded dark gray in the pulse profile). Top right panel is
	the phase-lag spectrum from the secondary peak (shaded light gray in the pulse profile).}
  \label{fig:pha}
\end{figure*}


The second possibility is emission from different poles. \citet{sas12} have analysed the
pulse profiles of this source and found a $\sim$ 60$^\circ$ offset between 
the position of one of the poles and the antipodal position of the other pole. 
This non dipolar field structure could lead to different local fields at the scattering
regions of each pole, which in turn could cause the difference in energies of the lines 
formed in each of them. The variation in accretion rate onto one pole as compared
to the other can cause the variation in the equivalent widths as seen in bottom panel of
Fig.~\ref{fig:tre}.

We can get further indicators by
looking at the pulse resolved energy spectrum and pulse phase-lag spectrum.
We examined the energy dependant phase lag of the pulses as computed by 
\citet{fer11} for the observation with simultaneous \rxte{}/\suzaku{} data. We have 
computed the phase lags using the data from the higher time resolution \rxte{} {\it PCA} 
detector, in a manner similar to that of \citet{fer11}.
As seen in Figure~\ref{fig:pha}, the pulse has two distinct peaks. We obtain the phase 
lags by taking phases corresponding to the main peak at a reference energy and 
cross-correlating them against the same set of phase bins at other energies. We have performed 
this for both the main and secondary peaks,  with phase bins taken as shown in bottom 
panel of Fig.~\ref{fig:pha}. In \citet{fer11}, the authors report that the most negative 
phase shifts occur at energies near the CRSF energies. This has been attributed 
to a change in beam pattern at the CRSF energies, with the photons at the CRSF band 
lying along a pencil beam whereas the rest of the emission is dominated by a fan beam.
 
In our analysis, we found that main peak's phase-lag spectrum had a similar set of
minima at $\sim$ 11\kev, $\sim$ 23\kev and $\sim$ 39\kev. However, the secondary
peak's phase-lag spectrum had minima at $\sim$ 16\kev and $\sim$ 30\kev. { The 
low count rate and increasing errors for energies above 45 \kev made it difficult
for us to find these dips at higher energies. }
The phase bins for performing the cross-correlation were chosen in such a manner
that the pulse profile at the reference energy band (taken as the band between
8.17\kev to 10.63\kev, similar to \citet{fer11}) would have a prominent peak
like structure in these phase bins. If we refer to the decomposed pulse profiles
in Fig. 8 of \citet{sas12} for the decay phase of the outburst, we see that
the two peaks in our overall pulse profile roughly correspond to the emission 
from the two poles. This lends support to the two pole possibility.
However, we cannot rule out the fan beam / pencil beam as we have not examined the
possible pulse profile and phase lags caused by such a possibility. Modelling these ``wavy''
phase lags, as attempted before \citep{fer11,sch14} could help resolve between these two 
possibilities. Pulse phase resolved
spectra could have given additional indicators. However, the large time resolution of
the \suzaku{} {\it XIS} data (of 2s) prevented us from getting the spectra at small phase
bins of the pulsar spinning at 3.6s.

One way to confirm our hypothesis would be to use a single wide-band large area detector
to make similar observations during the next outburst of this source. The 
constraint of using multiple instruments to bridge over the energy region having 
CRSFs and the region having the blackbody spectrum makes it very important 
to have these instruments cross-calibrated properly. While the cross-normalisation 
constants 
that we use compare favourably with the calibration carried out by \citet{tsu11}
for all instruments except \suzaku{} {\it PIN}-{\it XIS}, we do note that uncertainties in 
instrument cross-normalisations lead to uncertainties in the computed flux
values and line equivalent widths estimated. Having fewer number of instruments to cover
the range would then be ideal.
For example, five to six snapshot observations over different luminosities 
using the {\it XMM Newton} and {\it NuSTAR} telescopes would definitely give a higher
signal to noise data and lesser uncertainty in order to confirm or reject
our hypothesis. Another way to do this would be the possible construction of a polarization
spectrum. It is a well known fact that cyclotron resonant scattering has 
highly enhanced cross-sections for incident light polarized in the direction 
parallel to the local magnetic field, versus its cross-section for light polarized
perpendicular to it. This enhancement occurs at the resonant scattering
energy and drops off at other energies \citep[see][and references there-in]{bec07}. 
So, potentially a measurement of
polarization in small energy bands can pin-point the energy range over
which resonant scattering occurs, and effectively delineate the presence
of actual CRSFs from those resulting due to incorrect spectral modelling.
Future observations by detectors proposed for measuring the polarization in
different energy bands \citep{pau10,hay14} will definitely help improve our
understanding of this problem.

\subsection{Conclusion}
In this paper, we demonstrate the utility of having wide-band high signal to noise data
by making use of \suzaku{} and \rxte{} satellite data. Using this, we point out the presence of a 
blackbody component. We also note the possible indications of two sets
of CRSFs in this source.
We note that having two sets of lines at 11 \kev and 15 \kev explains the reason for the observed 
anti-correlation in the fundamental CRSF energy with source luminosity. It additionally explains the 
reason for this source being the only known accretion powered pulsar to show 5
harmonics of the fundamental cyclotron line. 
Data from large area wide-band telescopes like {\it NuSTAR} and {\it XMM} or from telescopes which can give a polarization
spectrum can potentially help solve this long open problem.

\section*{Acknowledgements}
This research has made use of data and software provided by the High Energy Astrophysics Science Archive Research Center (HEASARC), 
which is a service of the Astrophysics Science Division at NASA/GSFC and the High Energy Astrophysics Division of the 
Smithsonian Astrophysical Observatory. In addition, it made use of software from INTEGRAL Science Data Centre (ISDC). The
data themselves were from obtained from \suzaku{} (TOO proposer K. Pottschmidt), \rxte{} (TOO proposer K. Pottschmidt), 
\swift{} (TOO proposer S. M{\"u}ller) and \integral{} (TOO proposer S. Tsygankov) observatories. We acknowledge these
facilities and the proposers for making the observations and providing the data in the archives. We thank Prof. Carlo
Ferrigno for some useful pointers regarding the continuum at low energies. We acknowledge the referee's suggestions and
comments in helping us firm up the statistical anaylsis for line significance tests.
NI would like to thank Dr. Anil Agarwal, 
GD, SAG, Mrs. Valaramathi N, DD, CDA  and Dr. M. Annadurai, Director, ISAC for continuous support to carry out this research.

\appendix
\section{Fitting the continuum blackbody}\label{ap:bb}
  The blackbody component that we use for the continuum modelling does
  not consider effects of compton scattering of this component from the
  accreting plasma. \citet{far12} introduced an {\tt XSPEC} model for
  implementing this. The results of using this comptonized blackbody are
  detailed in this section. 

  There are nine parameters required to describe the comptonized blackbody
  against two required for the simple blackbody. For getting the model to fit, 
  we had to freeze the plasma parameters to the values obtained by previous 
  attempts to model this source using such a bulk comptonization of seed photons.
  \footnote{http://adsabs.harvard.edu/abs/2013A\%26A..553A.103F}.
  This model fits for the plasma electron temperature (kT${\rm_e}$  = 1.3 \kev)
  and optical depth ($\tau$ = 0.41), velocity profile of the accreting plasma (two parameters, $\eta$ = 0.5, $\beta$ = 0.22),
  radius of accretion column ($r_0$ = 0.1) and albedo percentage ($A$ = 1) from the NS surface\citep[see Table 1 of][]{far12}. 
  The remaining free parameters are the blackbody temperature and the model normalization. The results of 
  such a fit are summarized in Table \ref{tab:bbap}. As seen on comparing these results 
  with Table \ref{tab:bb}, we get blackbody seed temperatures to be  
  similar to the model with a simple blackbody. However, the radius of the blackbody drops to
  more reasonable values, although with larger errors. This gives a strong indication that a blackbody 
  component is indeed required to model the spectrum.

  \begin{table}
  \centering
  \caption{Blackbody temperature and radius for wide-band observations.}
  \begin{minipage}{0.5\textwidth}
  \begin{tabular}{lccc}
	\hline
	{Instrument} & {Day (MJD)} &  kT (\kev) & {Radius (kms)}\footnote{calculated from the normalization parameter} \\
	\hline
	\swift{} / \integral{} & 55736.34 & $1.19^{-0.26}_{+0.12}$ & $6.67^{-2.96}_{+3.28}$ \\
	\suzaku{} / \rxte{} & 55750.82 & $0.76^{-0.01}_{+0.01}$ & $6.21^{-1.15}_{+1.59}$ \\ 
	\hline
  \end{tabular}
  \end{minipage}
  \label{tab:bbap}
\end{table}

\bibliographystyle{mn2e}
\bibliography{ref}

\label{lastpage}
\end{document}